\documentclass[jcp,aip,onecolumn,groupedaddress,floats,showpacs,final]{revtex4-1}
\usepackage{graphicx}
\usepackage{bm}
\usepackage{color}
\usepackage{multirow}
\definecolor{blue}{rgb}{0.3,0.3,0.9}
\usepackage{amsmath}

\begin{document}

\title{Minimum Energy Pathways via Quantum Monte Carlo}

\author {S. Saccani$^1$, C. Filippi$^2$, S. Moroni$^{1}$}

\affiliation {$^1$SISSA Scuola Internazionale Superiore di Studi Avanzati and
        DEMOCRITOS National Simulation Center,
          Istituto Officina dei Materiali del CNR
         Via Bonomea 265, I-34136, Trieste, Italy\\
$^2$MESA+ Institute for Nanotechnology, University of Twente, P.O. Box 217, 7500 AE Enschede, The Netherlands}
\date{\today}
\begin{abstract}
We perform quantum Monte Carlo (QMC) calculations to determine minimum energy pathways of simple chemical reactions, and compare the computed geometries and reaction barriers with those obtained with density functional theory (DFT) and quantum chemistry methods. We find that QMC performs in general significantly better than DFT, being also able to treat cases in which DFT is inaccurate or even unable to locate the transition state. Since the wave function form employed here is particularly simple and can be transferred to larger systems, we suggest that a QMC approach is both viable and useful for reactions difficult to address by DFT and system sizes too large for high level quantum chemistry methods. 
\end{abstract}
%
%
\maketitle
\section{Introduction.}
Determining minimum energy pathways of reactions is of fundamental importance in scientific and
technological applications. The knowledge of barrier heights is key to the prediction 
of catalytic properties of materials since it enables the use of transition state theory (TST) 
to determine reaction rates~\cite{Eyring_TST,Wigner_TST,Keck_TST}. Locating efficiently the 
transition state on a potential energy surface (PES) is in fact a popular subject in 
computational physics and, to this aim, a variety of algorithms have been developed such as 
the shallowest ascent, synchronous transit, and nudged elastic band (NEB) approaches~\cite{Dykstra}. 
All these techniques ultimately rely on a method to determine the energy of a given atomic 
configuration and/or its derivatives. 

If we restrict ourself to quantum simulations, the most used
approaches are density functional theory (DFT) or highly-correlated quantum chemical methods, that is, 
wave function post-Hartree-Fock techniques such as the coupled cluster single-double and perturbative triple 
approach [CCSD(T)], which is generally considered the ``gold standard'' in quantum chemistry. Many of these
wave function methods are variational (though coupled cluster methods are not) and in principle offer a
systematic route to converge toward the exact energy, even though the increasing computational cost and the slow 
convergence severely limits this possibility. Their main drawback is that all these approaches implicitly 
or explicitly rely on expanding the wave function in Slater determinants and, therefore, require large amount of computer 
memory and have a poor size scaling ($N^7$ for CCSD(T), $N$ being the number of electrons), limiting their range 
of applicability to small systems.

Consequently, for larger systems, DFT remains the method of choice due to its much more favorable 
computational cost (scaling from $N^2$ to $N^4$).  Even though continuous progress in the field has
lead to the development of more precise and sophisticated DFT functionals, the situation is still far from satisfactory
if one aims at high accuracy~\cite{Zhang,Brittain}. For example, it is well known that popular functionals such as 
B3LYP~\cite{Lee_B3LYP,Becke_B3LYP,Stephens_B3LYP} often lead to poor transition state geometries and barrier heights~\cite{Nguyen,Zhao_NHTBH38/04_1}. 
Since DFT methods are not variational and do not offer a systematic way to improve their estimates, one has to resort 
to different approaches if better accuracy is needed.

Alternatively, one can employ quantum Monte Carlo (QMC) methods, such as variational
(VMC) and diffusion (DMC) Monte Carlo. These well-established ab-initio techniques take
advantage of Monte Carlo integration over the full Hilbert space. In particular, VMC is a stochastic way of calculating
expectation values of a complex trial wave function, which can be variationally optimized. DMC provides 
instead, with a higher computational cost,
a stochastic ground-state solution to the full Schr\"odinger equation, given a fixed nodal surface 
(using the fixed-node
approximation in order to avoid the notorious fermion sign problem). Because integrations
are performed in the full Hilbert space, one can make use of non separable wave functions, with the
explicit electron--electron correlation encoded in a so-called Jastrow factor. This allows for noteworthy accuracy already
using a simple and non memory-intensive single determinant Slater-Jastrow wave function. Although
considerably more expensive than DFT methods (scaling as $N^3$ with a much
larger prefactor), DMC generally offers better accuracy with respect to DFT. Furthermore, QMC methods 
possess a variational principle, which is a useful feature when one has to evaluate energy differences
as in TST. From a computational point of view, QMC codes can be made to scale linearly with the
number of cores and are not particularly memory demanding, making them suitable for today's
massively parallel supercomputers. Finally, QMC methods offer in principle the possibility
to push the calculation up to a desired accuracy by employing wave functions of increasing complexity
(although, from a practical point of view, one is likely to adopt simple wave functions for 
intermediate-to-large sized systems due to the increased computational cost of multi-determinant
wave functions).

In previous QMC studies of reaction barriers the geometries have been
taken either from DFT, or from constrained geometry optimization along
an assumed reaction coordinate \cite{Grossman1997,Barborini2012,Filippi2002}.
In this paper, we present nudged elastic band and climbing image
calculations~\cite{Mills_NEB,Henkelman_Climb_image}, where the geometry optimization of all the NEB 
images is done fully at the QMC level. For some representative challenging reactions from the NHTBH38/04 
database~\cite{Zhao_NHTBH38/04_1,Zhao_NHTBH38/04_2} and for a hydrogen transfer reaction~\cite{Peterson_OHH},
we determine transition state geometries and forward-reverse barrier heights within VMC and DMC, and compare 
our results against several current DFT functionals and other wave function methods. We demonstrate that VMC 
is able to locate reaction geometries with higher accuracy than DFT, while DMC outperforms DFT in 
evaluating barrier heights. Thus a sensible strategy, in terms of accuracy versus computational cost, is to calculate 
DMC barrier heights on VMC geometries.

\section{Methodology.}
The computation of the reactant and product geometries, the NEB calculations, and
the saddle-point location through the climbing-image method are all optimization procedures over the
total energy, although with different constraints. In particular, we use here the Newton optimization method,
based on the knowledge of the first and second derivatives of the energy. For the reactions reported here, 
these derivatives are from QMC calculations based on correlated sampling but, for larger systems, analytic
calculation of QMC forces is possible and advisable in order to reduce computational effort~\cite{Sorella_AlgDiff}.
Unless otherwise stated, we employ a single determinant Slater-Jastrow wave function:
\begin{eqnarray}\label{Wavefunction}
\Psi(r,R)=D^{\uparrow}(\phi^\uparrow,r^\uparrow,R) D^{\downarrow}(\phi^\downarrow,r^\downarrow,R) J(r,R),
\end{eqnarray}
where $\{r\}$ and $\{R\}$ denote the electronic and nuclear positions, respectively. 
The Jastrow factor, $J(r,R)$, explicitly depends on the inter-particle coordinates, and includes 
electron-electron, electron-nucleus, and electron-electron-nucleus correlation 
terms~\cite{Jastrow_filippi_umrigar}. The Slater determinants, $D^\uparrow$ and
$D^\downarrow$, are constructed from the sets of molecular orbitals $\{\phi^\uparrow\}$ and 
$\{\phi^\downarrow\}$ for the up- and down-spin electrons, respectively.
We employ scalar-relativistic energy-consistent Hartree-Fock pseudopotentials specifically
constructed for QMC calculations and expand the molecular orbitals on the corresponding 
cc-pVDZ basis set~\cite{burkatzki_1,burkatzki_2,H_pseudo}. These pseudopotentials have been 
extensively benchmarked, and their reliability has been recently supported by a DMC computation 
of atomization energies to near-chemical accuracy~\cite{Petruzielo}. 
The pseudopotentials are treated beyond the locality approximation~\cite{Casula2006}.

Our choice of such minimal wavefunction and basis set is intentional since we want to maximize the scalability 
of our approach to systems larger than the ones considered here. Our interest here is not to challenge 
quantum chemistry methods for small systems but, rather, to devise a strategy that has a more extended range 
of applicability while preserving a notable accuracy. While most DMC calculations found in literature use molecular
orbitals computed with some other electronic structure method~\cite{Williamson,Wagner1,Batista,Sola,Wagner2,Driver,Kent},
most often DFT, a key feature of our approach is that it is fully consistent since, at each iteration
step in our geometric optimization, we perform a QMC optimization of all wave function parameters.
This is done in order to guarantee consistency between the forces and the PES
(see below) as well as to improve the results in terms of the absolute energy. The wave function
optimization is performed at VMC level and details of the procedure are described elsewhere~\cite{Umrigar_optimization}. 
It is found that this optimization procedure only approximately doubles
the computer time needed to perform the calculations, while significantly lowering the expectation
value of the energy.  

The QMC calculations are performed with a modified version of the CHAMP program~\cite{CHAMP}. 
Since forces calculated in QMC possess a statistical uncertainty, strictly
speaking the optimization procedure via the Newton method never converges. Therefore, the equilibrium positions of the images along the nudged elastic band are obtained by averaging over several iterations after all quantities vary only by statistical fluctuations around a stationary value.  
We use a time step of 0.01 a.u. in the DMC calculations.
The DFT and Hartree-Fock (HF) all-electron calculations are performed with
the GAMESS package \cite{Schmidt_gamess}, using Dunning-type Correlation Consistent triple-zeta basis sets, augmented with a set of diffuse function (aug-cc-pVTZ). The QMC calculations employ instead scalar-relativistic pseudopotentials. For the light elements considered here, these relativistic effects are small and will not affect our results.

%
\begin{table*}
\caption{\label{Table_1} Barrier heights and RMS of geometric deviations, $\text{H + F}_2 \rightarrow \text{HF + F}$ reaction, for reactants (React), products (Prod) and transition states (TS). We denote with $V_f/V_r$ the forward/reverse reaction barrier heights (BHs). The RMS is calculated over the deviation of the interatomic distances of all the atoms from the best estimate geometry.}
\tabcolsep=0.05cm
\footnotesize
\begin{tabular}{| c | c c c c c c | c c c c c c c |}
\hline
\scriptsize{$\text{H + F}_2 \rightarrow \text{HF + F}$} & & \scriptsize{BE} & \scriptsize{VMC} & \scriptsize{VMC CAS} & \scriptsize{DMC} & \scriptsize{DMC CAS} & \scriptsize{HF} & \scriptsize{LSDA} & \scriptsize{BLYP} & \scriptsize{B3LYP} & \scriptsize{PBE} & \scriptsize{PBE0} & \scriptsize{M06} \\
\hline
\multirow{2}{*}{\parbox{23mm}{BHs \\ (Kcal/mol)}}                      & $V_f$ & $2.27$   & $6 \pm 1$   & $1.3 \pm 0.2 $  & $2.2 \pm 0.5$    & $1.4 \pm 0.3$& 0.0   & 0.0  & 0.0  & 0.0  &  0.0  & 0.0   & 2.7   \\
                                                                       & $V_r$ & $106.18$ & $126 \pm 1$ & $112.2 \pm 0.6 $& $114 \pm 1$    & $105.4 \pm 0.7$& 127.3 & 83.2 & 90.9 & 100.9 & 87.9  & 100.0 & 107.8 \\
\hline
\multirow{3}{*}{\parbox{23mm}{RMS deviation \\ (\AA)}}                 & React& & 0.008 & & 0.007 & & 0.067 & 0.010 & 0.037 & 0.002 & 0.018 & 0.019 & 0.020 \\
                                                                       & Prod & & 0.002 & & 0.008 & & 0.016 & 0.018 & 0.020 & 0.009 & 0.017 & 0.004 & 0.001 \\
                                                                       & TS   & & 0.028 & & 0.013 & & -     & -     & -     & -     & -     &   -   & 0.216 \\
\hline
\end{tabular}
\end{table*}

\section{Results.}
We select four challenging reactions from the NHTBH38/04 database~\cite{Zhao_NHTBH38/04_1}
plus one hydrogen transfer reaction. As best estimates, we use the atomic geometries
for the initial, final, and transition states reported in the database and computed through a quadratic 
configuration interaction with single and double excitations (QCISD) optimization. For these geometries, 
the barrier heights estimated with the W1 method (a complete basis set extrapolation over CCSD(T)) are also available. 
The reference data for $\text{H} + \text{OH} \rightarrow \text{H}_2 + \text{O} $ are from ext-CAS+1+2+Q 
calculations \cite{Peterson_OHH}.

We initially focus on the $\text{H + F}_2 \rightarrow \text{HF + F}$ reaction and collect the DFT and
QMC data in Table \ref{Table_1}. To measure how much the geometries differ from the best estimates, 
we calculate the RMS deviations of the interatomic distance among all atoms in the initial/final/transition state 
configurations with respect to the corresponding best-estimate geometries.
In the Table, a forward barrier ($V_f$) of zero means that the 
DFT functional finds no transition state (i.e.\ the reactants are unstable) with the reverse barrier being 
the reactant-product energy difference. Most DFT functionals fail in finding any transition state for this 
reaction, including the hybrid functionals PBE0 \cite{Adamo_PBE0} and B3LYP, while M06 \cite{Zhao_M06} retrieves 
a saddle point but with large deviations over the best estimate transition state geometry. In Table \ref{Table_1}, we also
report the initial/final/transition state geometries computed via VMC forces, where the uncertainty on the interatomic bonds 
due to the statistical noise on the forces is about 0.002 \AA. These geometries come from a fully VMC NEB and climbing image calculations. VMC is able
to retrieve even at the single-determinant level the initial, the final and, especially, the transition state geometry with much better accuracy than DFT. 
It is interesting to notice that, for geometrical data, VMC also performs better than the hybrid-meta M06 functional which is constructed
to fit the barrier heights calculated on the best-estimate geometries from the database~\cite{Zhao_M06}. Clearly, this 
fitting procedure does not always guarantee that the actual transition state retrieved by the functional is near the best-estimate
one. For testing purposes we also have located the saddle point using non optimized orbitals from the M06 functional and a fixed Jastrow; we find that the result is substantially worse, with a RMS deviation from the best estimate of 0.149 \AA. This result confirms that at least in some cases the wavefunction optimization is needed to obtain accurate values. 

Although the VMC geometries are rather accurate, the predicted reverse energy barrier ($V_r$) is markedly 
overestimated. Performing a DMC calculation on the VMC geometries retrieves better energy estimates, but the 
error on the reverse barrier of about 8 Kcal/mol is still quite significant.
We find that, to improve this energy barrier, it is not useful to reoptimize the geometries via DMC forces: 
the use of DMC forces does not alter significantly the geometries (within a statistical uncertainty on 
the transition state geometry of about 0.008 \AA) and, consequently, the estimated barrier heights either. In order to improve over
these values, it is possible to take advantage of the variational principle available in QMC and to
resort to the use of multi-determinantal wave function. In this case, we employ a simple complete active
space (CAS) wave function correlating three electrons in the three active orbitals relevant for the reaction 
and recompute the energy barriers over the VMC geometries obtained at the single-determinant level. The
resulting VMC barrier heights are improved and the DMC values become very similar to the best estimates.
Although the use of CAS wave functions is not readily applicable to larger systems due to their exponential scaling with system size, there exist
scalable techniques to improve over single-determinant wave functions through the design of accurate multi-determinantal size-extensive and
linear-scaling \cite{Fracchia} or backflow wave functions \cite{holzmann,lopez_rios}.

\begin{figure}
\includegraphics[scale=1.0]{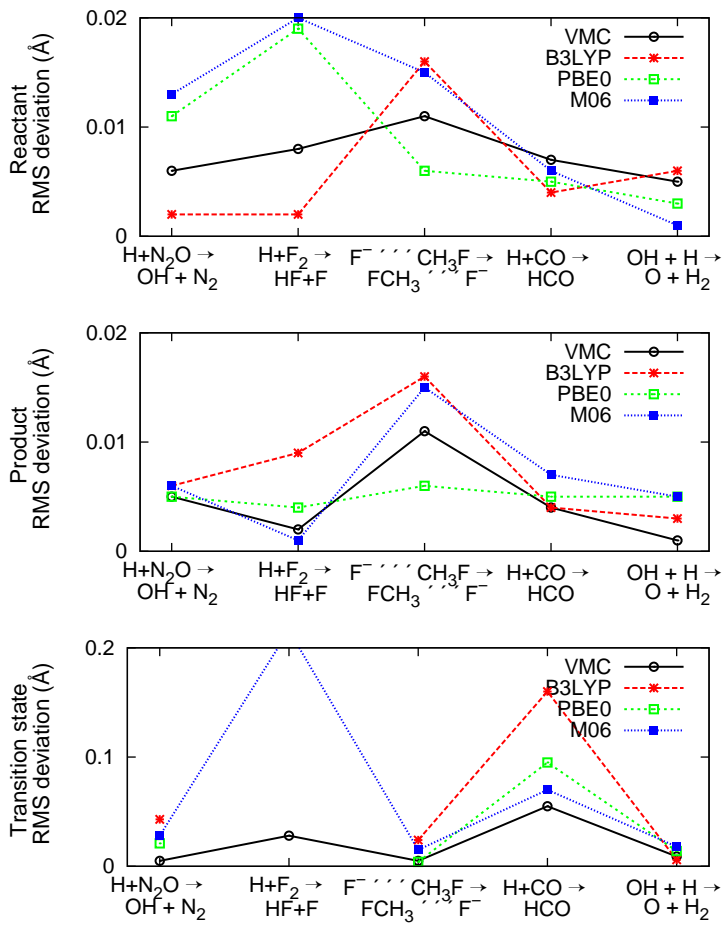}
\caption{\label{Fig_1} RMS of deviation of interatomic distances from the QCISD geometries (\AA). Distances are calculated among all atoms involved in the reactions.}
\end{figure}

We now return to the simple Slater-Jastrow wave function, and consider the other reactions.
In Figure~\ref{Fig_1}, we compare the difference between the geometries obtained by VMC and various DFT
functionals with respect to the reference ones obtained with QCISD, using the same measuring criterion as
in Table \ref{Table_1}. The functionals reported in this figure are all hybrid or meta-hybrid and are 
the ones returning the smallest geometric/energy deviation among the ones we tested on this set of reactions, 
which are the same ones listed in Table \ref{Table_1}. The generalized-gradient-approximation functionals used in the
overwhelming majority of DFT applications of CI-NEB, entail significantly larger deviations. For example, the deviation of PBE is
two to four times larger than that of PBE0 for transition state geometries reported here, and the RMS barrier heights deviation of PBE on this set is more than twice
than that of PBE0.
For equilibrium geometries, VMC performs at the level of the hybrid functionals. For the transition state, it
typically returns more accurate geometries, often performing much better than DFTs. Notwithstanding
that the M06 is actually fitted to reproduce barrier heights for the NHTBH38/04 reactions, VMC still performs
better than this functional in evaluating transition state geometries. This may be again due to the M06 parametrization
procedure, which does not guarantee the accuracy of the saddle point located by the functional. We do
not recalculate the geometries employing DMC forces because, from the test performed, DMC
forces improves VMC transition state geometries only slightly, as these are already notably accurate. Furthermore,
DMC geometry corrections are barely reflected in the calculation of the barrier heights, as these energies are second 
order in the deviation from the actual equilibrium points. For these reasons, we speculate that the
calculations of geometries at VMC level and of barrier heights at DMC level is the most sensible choice regarding
the trade-off between accuracy and computational cost.

\begin{figure}
\includegraphics[scale=1.0]{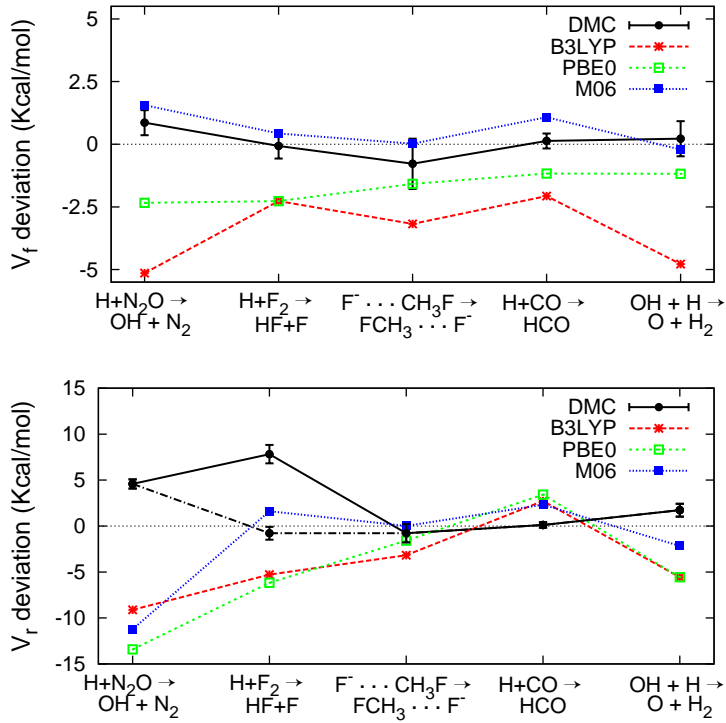}
\caption{\label{Fig_2} Forward ($V_f$) and reverse ($V_r$) barrier heights deviation from best estimates calculated by QMC and  DFT  Methods (Kcal/mol). The dash-dotted line represents the DMC values obtained by employing the CAS wavefunction for the $\text{H + F}_2 \rightarrow \text{HF + F}$ instead of the single determinant one (full line), see text.}
\end{figure}

In Figure \ref{Fig_2}, we report the forward and reverse reaction barriers. While 
VMC (not shown) performs less accurately than hybrid functionals, 
DMC calculated on the VMC geometries significantly improves the barrier heights of all reactions upon the VMC values,
and performs at the level of the hybrid DFT approaches. In particular, our QMC procedure is more accurate
than the hybrid functionals B3LYP and PBE0, while the only functional performing on average at the same level 
or slightly better is M06, despite these barriers being calculated on transition state geometries worse than VMC ones. Moreover M06 is actually fitted to reproduce precisely the NHTBH38/04 barrier heights, and there is no guarantee that, on a different set of reactions, it would still perform as well as QMC. Note that single-determinant DMC performs better than all DFT approaches, included M06, even in the reaction $\text{H} + \text{N}_2\text{O} \rightarrow \text{OH} + \text{N}_2$, which is known to be strongly multi-determinantal in character. Overall, QMC gets both the geometry and the energetics accurately, offering a parameter-free, more balanced description of reactants, products and transition states than all DFT schemes considered here.

\section{Remarks on QMC Forces calculation.}
We compute here the VMC and DMC energy derivatives with
correlated sampling. While the procedure is relatively straightforward in VMC, the calculation
of forces in DMC is more involved and approximations are generally used. For details about the DMC algorithm,
we refer the reader to Ref.~\cite{filippi_umrigar_forces}. In both VMC and DMC, one should know the derivatives
of the energy with respect to the wave function parameters and of the parameters with respect to the nuclear
coordinates in order to calculate the forces correctly. In VMC, we avoid this problem by fully optimizing 
the wave function so that all the derivatives with respect to the parameters are zero. In DMC, the
bias in the forces due to wave function optimization at VMC level has been found negligible in a few test
cases.

\section{Conclusions.}
We investigated the possibility of performing full-QMC reconstruction of minimum
energy pathways of chemical reactions. Full optimization of the wave function parameters is carried out
during each iteration, so the employed technique is internally fully consistent. Geometric optimization of
the minimum energy pathway and of the transition state is done at VMC level, with the obtained geometries
being more accurate than DFT ones, especially for transition states. It also demonstrates the ability
to correctly locate the transition state in cases in which DFT fails in returning accurate geometries. At DMC level,
the method displays very good performance in evaluating barrier reaction heights, comparing favorably
even against hybrid functionals. Therefore, our approach of calculating the geometries at the VMC level and
the barrier heights at the DMC level is most effective as far as performance over computational cost is concerned:
calculating DMC geometries is very expensive and, in the tested cases, it does not improve 
significantly the estimates, while calculating DMC energies over VMC geometries is much cheaper and 
still retrieves good results.
Since the employed wave function is of the simple Slater-Jastrow type, this technique is scalable to
larger systems. Our results indicate that, for intermediate-sized system reactions where quantum chemistry 
methods are not computationally viable, the QMC approach may be the most accurate technique currently 
available.

\subsection*{Acknowledgements.}
We acknowledge Dr. O. Valsson for assistance in the usage of CHAMP and GAMESS packages, and Prof.
S. Baroni for helpful discussions.

\end{document}